# Strong inelastic scattering of slow electrons by optical near fields of small nanoparticles

**Germann Hergert[1], Andreas Wöste[1], Petra Groß[1], and Christoph Lienau[1,2]**

[1] Carl von Ossietzky Universität, Institut für Physik and Center of Interface Science, 26129 Oldenburg, Germany
[2] Carl von Ossietzky Universität, Forschungszentrum Neurosensorik, 26129 Oldenburg, Germany

E-mail: christoph.lienau@uol.de

**Abstract**

The interaction of swift, free-space electrons with confined optical near fields has recently sparked much interest. It enables a new type of photon-induced near-field electron microscopy, mapping local optical near fields around nanoparticles with exquisite spatial and spectral resolution and lies at the heart of quantum state manipulation and attosecond pulse shaping of free electrons. The corresponding interaction of optical near fields with slow electrons has achieved much less attention, even though the lower electron velocity may enhance electron-near-field coupling for small nanoparticles. A first-principle theoretical study of such interactions has been reported very recently [N. Talebi, Phys. Rev. Lett. 125, 080401 (2020)]. Building up on this work, we investigate, both analytically and numerically, the inelastic scattering of slow electrons by near fields of small nanostructures. For weak fields, this results in distinct angular diffraction patterns that represent, to first order, the Fourier transform of the transverse variation of the scalar near-field potential along the direction perpendicular to the electron propagation. For stronger fields, scattering by the near-field component along the electron trajectory results in a break-up of the energy spectrum into multiple photon orders. Their angular diffraction patterns are given by integer powers of the Fourier transform of the transverse potential variation and are shifting in phase with photon order. Our analytical model offers an efficient approach for studying the effects of electron kinetic energy, near field shape and strength on the diffraction and thus may facilitate the experimental observation of these phenomena by, e.g., ultrafast low-energy point-projection microscopy or related techniques. This could provide simultaneous access to different vectorial components of the optical near fields of small nanoparticles.

## 1. Introduction

When free swift electrons pass an optically excited nanostructure at close distance, their wave function acquires a phase modulation. This phase modulation lies at the heart of photon induced optical near field microscopy (PINEM) [1-3] and resulted in the development of electron energy gain spectroscopy (EEGS) [4, 5]. These comparatively novel spectroscopic techniques enable local (transmission) electron spectroscopy with an energy resolution limited by the spectral width of the optical field rather than by the energy resolution of the electron beam. For sufficiently strong fields this phase modulation can be used to tailor the quantum state of free electron wave functions [6], opening up exciting new ways for the creation of attosecond electron pulse trains [7], or to directly measure the quantum state of nanolocalized optical fields [8].

Inducing such a phase modulation of the electron wave function is most efficient if phase matching between the localized field and the passing electron wave is satisfied: In this case, the optical wave vector component parallel to the propagation direction of the electron matches the ratio of optical frequency and electron velocity [3]. For swift electrons with velocities on the order of ${}^{2}/_{3}c_0$ this relation can be fulfilled, e.g., by letting the electrons pass evanescent fields at interfaces [9-11], in the vicinity of dielectric resonators [12, 13], or for optical near fields around nanostructures with dimensions below the optical wavelength [1, 2, 14]. Since the electron beam width employed in transmission electron microscopes (TEM) typically is on the order of only a few nm, the spatial variation of the optical field across the electron beam can be neglected and the phase modulation is described reasonably well in one-dimensional models [3]. Due to their high velocities, electrons in a TEM pass the optical field around particles with dimensions below 100 nm within less than an optical cycle. Since their transit time through the near field decreases even more with decreasing particle size, reaching a discernible phase modulation of swift electrons becomes increasingly demanding. Only recently PINEM was demonstrated for highly localized near fields of plasmonic nanostars [15] and most PINEM experiments have studied nanostructures with dimensions far above 10 nm [1, 16].

For small nanostructures, efficient electron near-field coupling can be improved by using sufficiently slow electrons. For such slow electrons, however, phase matching can only be reached in the near field of very small particles and, in fact, has not been demonstrated experimentally yet. Recent progress in low-energy electron microscopy brings such studies into reach. An especially promising realization lies in an ultrafast point-projection electron microscope (UPEM), where plasmonic nanofocussing is used to trigger photoemission from the apex of a metal tip, creating a free-standing source of low-energy electrons. A specific advantage of UPEM is its intrinsically high time resolution of currently ~20fs, reached in the absence of advanced compression schemes [17].

So far, the interaction between slow electrons and confined optical fields has not yet been studied in much detail. Recently, a first- principle description of such interactions showed that new phenomena arise, which are not observed for swift electrons [18]. Photon-order sidebands in the kinetic energy spectra, similar to those seen for swift electrons, are induced by the phase-matched longitudinal optical field component. In addition, the simulations show pronounced angular electron deflections with complex diffraction patterns.

Here, we analyze such slow-electron near-field couplings by presenting numerical as well as analytical solutions of the two-dimensional time-dependent Schrödinger equation. For electron wavepackets passing the confined dipolar fields of small nanostructures we observe quantized modulations of the electron momentum distribution in transverse direction. In some analogy to the Aharonov-Bohm effect [19], the resulting light-driven double-slit-like interference pattern is caused by a transversely-varying phase modulation of the electron wavepacket. The pattern reflects the Fourier transform of the transverse near-field component perpendicular to the propagation direction. The experimental investigation of these interferograms could pave the way towards a full vectorial characterization of optical near-field dynamics of individual nanostructures with few-femtosecond time resolution.

## 2. Methods

We model the propagation of a single-electron wavepacket $\psi(x,y,t)$ by solving the time-dependent Schrödinger equation in two dimensions

$$i\hbar\frac{d}{dt}\psi(x,y,t)=\hat{H}\psi(x,y,t) \ , \tag{1}$$

using the minimal coupling Hamiltonian

$$\hat{H}=\frac{1}{2m}\left(\hat{\mathbf{p}}-q\mathbf{A}\right)^2+q\Phi \ . \tag{2}$$

Here, $\hat{\mathbf{p}}=i\hbar\nabla$ is the momentum operator, $\mathbf{A}(x,y,t)$ the classical vector potential and $\Phi(x,y,t)$ the classical scalar potential. The electron mass and charge are $m$ and $q=-e$, respectively. In numerical solutions of equation (1), we use a linearly polarized, monochromatic plane-wave incident laser field with vector potential $\mathbf{A}_L(x,y,t)$. Its electric field, with spatially homogeneous amplitude $E_L$ optically excites a nanostructure and induces a local optical near field with the potential $\Phi_{NF}(x,y,t)$. For sufficiently slow electrons, this near field dominates the interaction with the electron, while the induced vector potential is negligible. In the analytic model described below, we neglect the interaction of $\mathbf{A}_L(x,y,t)$ with the electron due to the finite wavevector mismatch. The Hamiltonian thus reduces to $\hat{H}=\hat{\mathbf{p}}^2/\left(2m\right)+q\Phi_{NF}$.

We follow the approach introduced by Park et al. [3] for solving a one-dimensional Schrödinger equation model. We assume that the electron propagates in longitudinal *x*-direction with initial momentum $\mathbf{k_0}=k_0\mathbf{e_x}$ and separate its wavefunction into a product

$$\psi(x,y,t)=g(x-v_0t,y,t)\,\psi_0(x,y,t) \tag{3}$$

Here $g(x-v_0t,y,t)$ is the envelope moving with velocity $v_0=\hbar k_0/m$, and $\psi_0(x,y,t)=\exp(ik_0x-iE_0/\hbar t)$ is the carrier wave with initial electron kinetic energy $E_0$. Inserting (3) into (1) yields $i\hbar\dot{g}=-\hbar^2\Delta g/2m+q\Phi_{NF}g$. Neglecting wavepacket dispersion during the few-fs interaction time, the solution is given by

$$g(x',y,t)=g(x',y,t_0)\exp\left[-\frac{iq}{\hbar}\int_{t_0}^{t}\Phi_{NF}(x'+v_0t',y,t')dt'\right] \tag{4}$$

Here we introduced $x'=x-v_0t$ as the coordinate for the moving frame of reference of the envelope function $g$. In the integrand, we substitute $x''=x'+v_0t'$ and find that the envelope at time $t$ after interaction differs from that at $t_0$ only by a phase factor:

$$g(x',y,t)=g(x',y,t_0)\exp\left[-\frac{iq}{\hbar v_0}\int_{x''(t_0)}^{x''(t)}\Phi(x'',y,\frac{x''}{v_0}-\frac{x'}{v_0})dx''\right]=:g(x',y,t_0)e^{i\Delta\varphi(x',y)} \tag{5}$$

In contrast to the well-known 1D result [2, 3], the phase factor $\Delta\varphi(x',y)$ now depends on both spatial coordinates. In the following, we analyze the phase modulation that is acquired by a wave packet during its interaction with optically excited nanostructures.

For monochromatic excitation at optical frequency $\omega$, the near field potential $\Phi_{NF}$ can be written as:

$$\Phi_{NF}=\Phi_0(x,y)\cdot\cos(\omega t+\varphi_{NF}) \tag{6}$$

Inserting $\Phi_{NF}$ into equation (5), the phase modulation $\Delta\varphi$ is

$$\Delta\varphi=-\frac{q}{\hbar v_0}\int_{x''(t_0)}^{x''(t)}\Phi_0(x'',y)\cos(\Delta kx''-\Delta kx'+\varphi_{NF})dx'' \tag{7}$$

with the wavevector mismatch $\Delta k=\omega/v_0$. Using angle sum identities, equation (7) can be written as:

$$\Delta\varphi=I_1(y)\cos(\Delta kx')+I_2(y)\sin(\Delta kx') \tag{8}$$

The phase modulation thus is given by the sum of an even and an odd function multiplied with the coupling integrals $I_1(y)$ and $I_2(y)$, respectively:

$$I_1(y) = -\frac{q}{\hbar v_0} \int_{x''(t_0)}^{x''(t)} \Phi_0(x'', y)\cos(\Delta k x'' + \varphi_{NF})\, dx''$$
$$I_2(y) = -\frac{q}{\hbar v_0} \int_{x''(t_0)}^{x''(t)} \Phi_0(x'', y)\sin(\Delta k x'' + \varphi_{NF}) dx'' \qquad (9)$$

These coupling integrals depend on the transverse coordinate *y* and are given by Fourier components of $\Phi_0(x'', y)$ along the dimension $x"$ at frequency $\Delta k_x$. The *y*-dependence of the coupling constant is the essential difference between the two-dimensional and one-dimensional simulations.

## 3. Results and discussion

Using the derived formalism, we simulate the interaction of a single-electron wavepacket with an infinitely long, thin wire of dielectric function $\varepsilon$ and radius *R*. The electron is propagating in *x*-direction and the wire is oriented perpendicular to the simulation (*x*-*y*) plane. In quasi-static approximation the wire potential $\Phi_0$ for a linearly *y*-polarized excitation at field amplitude $E_L$ can be written as [20]:

$$\Phi_0 = \begin{cases} E_L y \left| \dfrac{\varepsilon - 1}{\varepsilon + 1} \right|, & r < R \\ E_L y \dfrac{R^2}{r^2} \cdot \left| \dfrac{\varepsilon - 1}{\varepsilon + 1} \right|, & r \geq R \end{cases} \qquad (10)$$

$$\varphi_{NF} = \mathrm{atan}\left( \frac{\varepsilon - 1}{\varepsilon + 1} \right) \qquad (11)$$

Here, $\varphi_{NF}$ is the phase retardation with respect to the incident laser field that is in induced by the complex-valued dielectric function $\varepsilon$. Since the vector potential $A_L$ has no direct consequence on the electron propagation and since the electron pulse duration is longer than one optical cycle, a variation in $\varphi_{NF}$ does not affect the result. In the following, we set this phase to zero. As a consequence, the coupling integral $I_2(y)$ in equation (9) vanishes due to the symmetry of $\Phi_0$. To mimic a carbon nanotube (CNT) that is transparent even for slow electrons, we take the dielectric function of a CNT film [21]. The resulting scalar potential of a 10-nm radius CNT at a wavelength of 2000 nm and for $E_L = 0.2\ \mathrm{V/nm}$ is seen in figure 1(a). Initially, we consider a slow electron wavepacket with $E_0 = 100\,e\mathrm{V}$ and longitudinal and transverse broadenings of 60 nm and 20 nm, respectively. The longitudinal broadening corresponds to a temporal spread of ~10 fs. The initial electron density at time $t_0$, $|\psi_i(x,y)|^2$, is displayed on the left side in figure 1(a). After $t - t_0 = 60\,\mathrm{fs}$ of propagation the electron wavepacket has passed the CNT, and a pronounced modulation of $|\psi_f(x,y)|^2$ due to its interaction with the optical near field emerges. The quiver motion of the electron creates an oscillatory bunching pattern along both, longitudinal and transverse directions. In momentum space this modulation leads to distinct peaks in the associated density distribution $|\tilde{\psi}_f(k_x,k_y)|^2$ (figure 1(b)) with well-defined spacing $\Delta k_x$ along the longitudinal direction and spacing $\Delta k_y$ along the transverse direction. The resulting diffraction pattern is quantitatively reproduced by a numerical solution of the 2D Schrödinger equation (figure 1(c)). These numerical simulations include the effect of the vector potential on the electron motion as well as the dispersion of the wavepacket during the near-field interaction.

For understanding the diffraction pattern, it is important that the near-field interaction couples the initial momentum state $\mathbf{k_0} = k_0\mathbf{e_x}$ of the incident electron to different momentum states on the free-electron dispersion relation. In principle, all final states are allowed that fulfil energy- and momentum conservation in this electron-near-field interaction. For weak driving fields, the change in the electron kinetic energy $\Delta E$ can be approximated as $\Delta E \propto \left(k_0\mathbf{e}_x \pm \Delta\mathbf{k}\right)^2 - k_0^2 \approx \pm 2k_0\Delta k_x$, since the transferred momentum $|\Delta\mathbf{k}|$ from the near field is much smaller than $k_0$. Hence, the near-field interaction causes a defined longitudinal momentum change $\Delta k_x = \omega / v_0$. The energy change is essentially independent of $\Delta k_y$ for slow as well as for swift electrons, and energy- and momentum conservation applies no selection rule for $\Delta k_y$. Thus, in principle all transverse momentum components $\Delta k_y$ of the near field can be transferred. The observation of well-defined peaks along $k_y$ in figures 1(b) and (c) therefore implies that only selected components $\Delta k_y$ are available in the near field. Effectively, the diffraction pattern thus provides, for sufficiently weak field amplitudes of the incident laser, the momentum components $\tilde{\Phi}_{NF}(k_x = \pm\Delta k_x, k_y)$ of the transverse Fourier transform of $\Phi_{NF}(x, y)$ at the wavevector $\Delta k_x$. The allowed first-order transitions are depicted as blue arrows in figure 1(d). Here the electron *k*-states that are populated by near-field scattering are depicted as blue circles and the solid black lines define states with constant kinetic energy, spaced by integer multiples *n* of the photon energy $\hbar\omega$. The peaks along $k_x$ appear at the positions of the well-known PINEM sidebands in the electron kinetic energy spectrum [3]. To elucidate the structure in *y*-direction we perform a Taylor expansion of $e^{i\Delta\varphi(x',y)}$ (equation (5)) and sort the terms by photon orders *n* of $\cos(n\Delta k_x x')$. The final wavefunction, after the near-field interaction, for a given order *n* then can be expressed as [22]:

$$\tilde{\psi}_f(k_x = n\Delta k_x + k_0, k_y) \propto \tilde{\psi}_i(k_x = k_0, k_y) \otimes FT\left\{\sum_{\ell=|n|}^{\infty} \frac{i^{2\ell}}{2^{2\ell}(\ell - |n|)!\ell!} I_1(y)^{2\ell-|n|}\right\} \tag{12}$$

Equation (12) represents a summation over all possible excitation paths in different powers of $I_1(y)$ that lead to the final states at $(k_x = k_0 + n\Delta k_x, k_y)$. Here the exponent $2\ell - |n|$ can be understood as the number of photons exchanged between electron and near field. The magnitude squared of $\tilde{\psi}_f$ provides the probability for occupying a certain final momentum state as a consequence of the multilevel Rabi oscillations [2] that are driven by multiple near-field-photon absorption and stimulated emission processes (figure 1(d)) [23]. For sufficiently weak fields the contribution of higher-order scattering terms with $\ell > |n|$ can be neglected and the above equation can be approximated as:

$$\begin{aligned} \tilde{\psi}_f(k_x = n\Delta k_x + k_0, k_y, t) &\propto \tilde{\psi}_i(k_x = k_0, k_y) \otimes FT\left\{I_1(y)^n\right\} \\ &\propto \tilde{\psi}_f(k_x = (n-1)\Delta k_x + k_0, k_y, t) \otimes \tilde{I}_1(k_y) \end{aligned} \tag{13}$$

This retains only the most direct scattering path between initial and final state and neglects all scattering paths that contain both absorption and stimulated emission. In this limit, the wavefunction along $k_y$ for a certain order *n* thus can be obtained by a convolution of the wavefunction at the previous order (*n-1*) and the Fourier transform $\tilde{I}_1(k_y)$ of the coupling integral $I_1(y)$.

The coupling integral for the case of the CNT is shown in figure 1(e). The magnitude of its Fourier transform, shown in the inset, reveals the peak splitting of ~ $2\Delta k_y$ that defines the peak positions in the diffraction pattern. Figures 1(f) and (g) show crosscuts of the final momentum distribution of the electron, $|\tilde{\psi}_f|^2$, along the coloured lines in figure 1(b). For a given $\Delta k_y$, the cross sections in figure 1(f) show dominant peaks that are separated by $2\Delta k_x$, since the final states differ by an even number of photon transitions. In addition, fainter peaks can be seen between these due to the finite transverse momentum spread of the initial wavepacket and the finite width of $\tilde{I}_1(k_y)$. Crosscuts along $k_y$ at different photon orders (solid lines in figure 1(g)) show peaks that are spaced by $2\Delta k_y$, for the same reason as in figure 1(f). For such comparatively weak driving fields, the pattern is reasonably well reproduced by the approximation in equation (13) (dashed lines).

It is evident that the interference patterns for cross sections at consecutive photon orders shift by $\Delta k_y$. This shift resembles the phase shift of the double-slit interference pattern in the Aharonov-Bohm effect that is induced by the local vector potential of a current-carrying solenoid. There, the electron passes on opposite sides of the solenoid, traveling either parallel or antiparallel to the vector potential. This introduces a transverse phase modulation on the electron wavefunction that leads to the fringe shift. In our case, the phase modulation is induced by traversing the scalar potential in the vicinity of the nanostructure (equation (5)). The near-field potential flips sign on the opposing sides of the nanowire (figure 1(a)), introducing a transverse phase modulation. The amplitude of the phase modulation scales linearly with laser field strength. The field-controlled change in $\Delta\varphi(x,y)$ induces the $\pi$- phase shifts of the diffraction pattern along $k_y$ between consecutive photon orders *n*. This light-driven phase modulation offers a conceptually novel approach towards coherent control of ultrafast electron diffraction by strong local optical near fields.

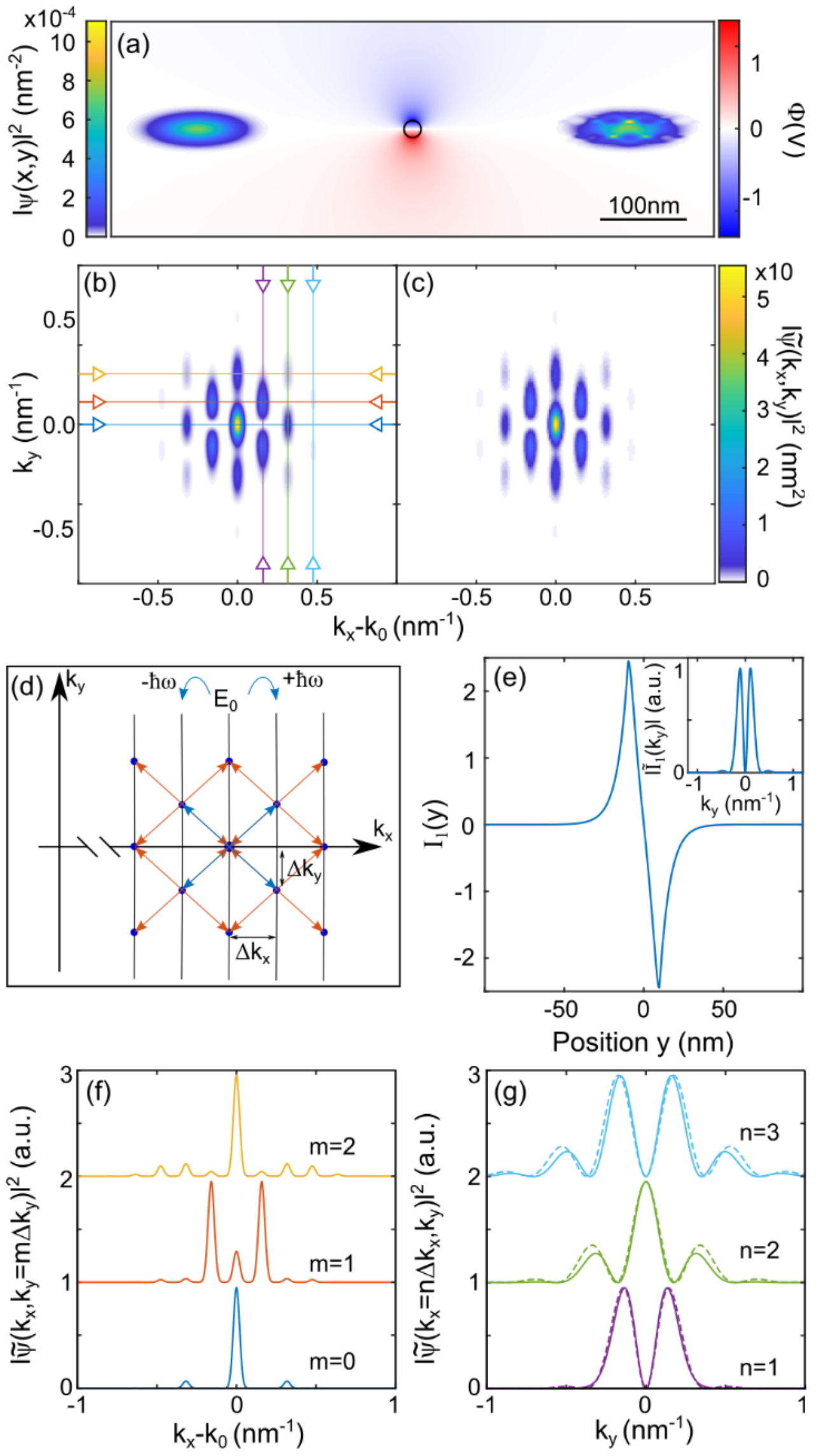

Figure 1. Low-energy electron wavepacket propagation through the near-field potential of a thin wire with a radius of 10 nm and a dielectric function matching that of a carbon nanotube. The wire is optically excited at 2000 nm with a linearly *y*-polarized field with an amplitude of 0.2 V/nm. The field enhancement at the surface of the wire is 1.5. (a) The density distribution of the incident 100-eV electron has a width of 60 nm along the propagation direction (*x*), corresponding to a 10-fs pulse, and 20 nm in transverse (*y*) direction. The phase modulation introduced by the near-field coupling results in a bunching of the electron density along both coordinates. (b,c) Final wavepacket density $\left|\tilde{\psi}_f\left(k_x,k_y\right)\right|^2$ in momentum space calculated analytically (b) and numerically (c), showing diffraction peaks with distinct spacings $\Delta k_x$ and $\Delta k_y$ along both momentum directions. (d) Allowed k-space

transitions for the near-field interaction. Vertical black lines represent electron states with energies spaced by the photon energy. (e) Calculated coupling integral $I_1(y)$ and its Fourier transform $\tilde{I}_1(k_y)$ (inset). (f,g) Crosscuts through $\left|\tilde{\psi}_f\left(k_x,k_y\right)\right|^2$ at multiples of the spacings $\Delta k_y$ (f) and $\Delta k_x$ (g).

In order to investigate how the coupling between electron and near field scales with the kinetic energy of the electron, we now perform the same calculation with different $E_0$. To obtain sufficiently high coupling for all considered energies, the incident field strength is increased to $E_L = 0.5V / nm$. The longitudinal spread of all wavepackets is adjusted to 500 nm, such that all examined electrons have a temporal spread exceeding one optical period. Exemplarily, the resulting final momentum distributions $\left|\tilde{\psi}_f\right|^2$ for four different energies are shown in figures 2(a)-(d). Since an increase in electron energy leads to a decrease in the longitudinal momentum change $\Delta k_x$ , the axes in the figures are scaled accordingly. For low-energy electrons, the increased field strength leads to significant contributions from multiple interfering excitation pathways, including both absorption and stimulated emission of photons. This results in a more complex diffraction pattern in figures 2(a), (b) than seen in figure 1(b). With increasing electron kinetic energy (figures 2(c), (d)) the phase mismatch increases, reducing the effective coupling strength. Figure 2€ shows crosscuts of $\left|\tilde{\psi}_f\right|^2$ at $k_y = 0$ along the $k_x$-direction as function of initial kinetic energy. Here, higher photon orders are only visible for an efficient electron-near-field coupling, which appears for the given CNT radius $R = 10\,\mathrm{nm}$ only between 100 eV and 2 keV. Additionally, a full depletion of the ground state is visible around 650 eV and 3.3 keV. Analogously, figure 2(f) shows crosscuts of $\left|\tilde{\psi}_f\right|^2$ along $k_y$ at $k_x = 0$. As for figure 1(e), occupation of higher $k_y$-states decreases with increasing electron energy. More importantly, the resulting angular deflection $\alpha = \mathrm{atan}\left(k_y / k_0\right)$ for a fixed $k_y$ increases for lower electron velocities, leading to angles of up to $\alpha \approx 1^\circ$ for electron energies around 100eV, which can be easily resolved in UPEM.

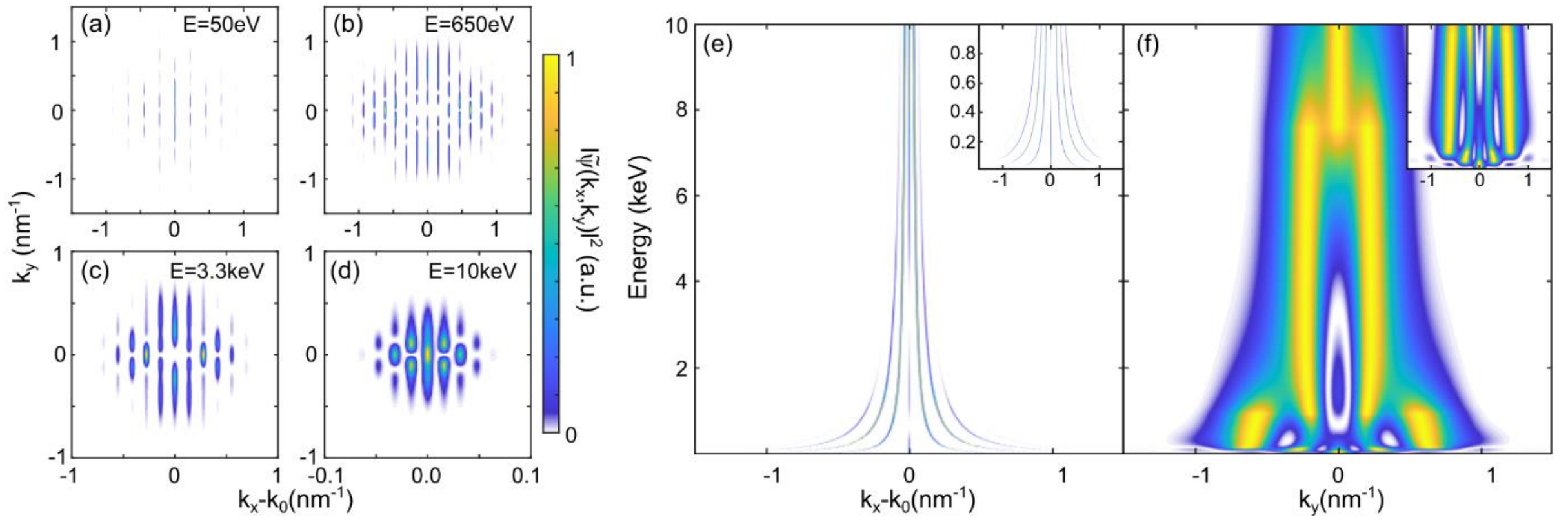

Figure 2. Effect of the electron kinetic energy $E_0$ on the near-field-induced diffraction by a 10-nm radius wire. The simulation parameters are chosen as in figure 1, except for a longitudinal width of 500 nm and an increased field amplitude of 0.5 V/nm. (a-d) Final momentum density $\left|\tilde{\psi}_f\left(k_x,k_y\right)\right|^2$ for electron energies between 50 eV and 10 keV, showing a maximum number of diffraction orders at 650 eV. (e, f) Energy dependence of the final momentum density for $k_y$ = 0 (e), showing a reduction in $\Delta k_x$ with $E_0$, and for $k_x$ = 0 (f), at constant $\Delta k_y$. An optimum interaction strength is reached at 100 eV due to the increased interaction time with the optical near field. For slower electrons, phase matching is no longer fulfilled.

In a next step we investigate the influence of the CNT radius on the scattering for an initial electron energy of 100 eV. For each radius, the transverse width of the electron wavepacket is set to 2R, to ensure that it passes the opposing sides of the CNT. Figures 3(a)-(d) show resulting diffraction patterns at four selected radii. It is evident that the coupling strength reduces with increasing radius. In figure 3(e) crosscuts of the diffraction pattern along $k_x$ at $k_y = 0$ are shown in dependence of the radius. Here, additionally to the overall decrease in coupling strength, peaks up to the sixth photon order can be seen in $k_x$ direction for different radii. The emerging and vanishing of the higher order peaks with increasing radius shows recurrent efficient coupling also for larger radii. Similar oscillations in coupling efficiency are also observed in the crosscuts along $k_y$ at $k_x = 0$ in figure 3(f), but with decreasing $\Delta k_y$ for increasing radius. This can be explained by the reduced amplitude of the Fourier components of $\tilde{\Phi}(k_x = \Delta k_x, k_y)$ at large $k_y$ values for increasing CNT radii. For 100-eV electrons and a laser wavelength of 2000 nm, we find an optimum coupling efficiency for a radius of 10 nm. Treating the electron as a classical point-particle this would correspond to a transit time through the near-field of exactly a half period of the optical cycle $2R = vT/2 = \pi / \Delta k_x \approx 20nm$ [18].

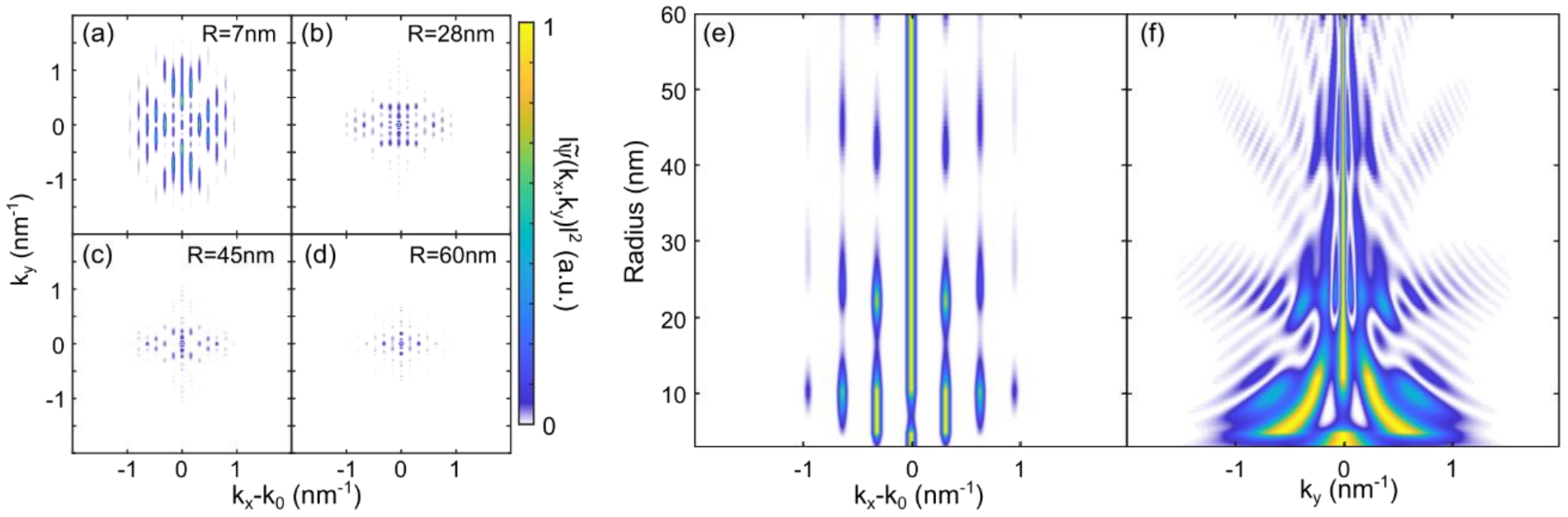

Figure 3. Effect of the wire radius R on the near-field-induced diffraction of a 100-eV electron wavepacket. The transverse width of the electron is adapted to 2R and the field amplitude $E_L$ is set to 0.5 V/nm, all other simulation parameters are chosen as in figure 1. (a-d) Final momentum density for wire radii between 7 nm and 60 nm. (e, f) Cross sections through the final momentum density along $k_x$ (e) for $k_y$ = 0 and $k_y$ (f) for $k_x$ = 0. For wire radii below 5 nm, the interaction time is too short to reach meaningful coupling strengths. Phase matching cannot be fulfilled with radii much above 10 nm, resulting in oscillations and an overall decrease in interaction strength with radius.

The simulations show how efficient coupling between the near field of an optically excited, nanometer-sized structure and low-kinetic-energy electrons can be achieved. However, wires with 10-nm radius that are transparent for slow electrons are not readily available experimentally. In figure 4 we extend these simulations to a realistic sample geometry, providing a similar near-field potential. We consider a nanoresonator, which is milled into a free-standing gold film with a thickness of 13 nm, shown in figure 4(a). The scalar potential for this geometry is modelled as the potential of two dipole distributions, separated by 23 nm in *y*-direction, in Lorenz gauge. The dipole distributions account for the finite sample size by convoluting a point dipole potential with a 2D-Gaussian function with 13-nm FWHM in both dimensions. The resulting potential $\Phi_{NF}(x, y)$ is displayed in figure 4(b), reproducing the potential of a plasmonic gap mode obtained by FDTD simulations reasonably well. The amplitude of the potential is chosen such that the maximum field strength inside the resonator is 0.5 V/nm, corresponding to an incident field strength of $E_L < 0.03\,\mathrm{V/nm}$ for typical field enhancements of ~20 for such resonators. For the simulation we assume electrons with 100-eV kinetic energy, propagating in the *x*-direction, perpendicular to the sample plane. The longitudinal spread is chosen to give a bandwidth-limited temporal spread of 20 fs. The transverse spread is limited to 5 nm to emulate the measureable signal of electrons passing the gap and impinging on a detector. The resulting diffraction pattern $\left|\tilde{\psi}_f\left(k_x, k_y\right)\right|^2$ resembles the results obtained for the wire geometry and is shown in figure 4(c). However, the spread in kinetic energy for such a bandwidth-limited electron is only 0.05 eV, which is magnitudes smaller than what is experimentally feasible. To resemble a possible experiment more closely the bandwidth is increased to 2 eV, while the temporal spread is kept at 20 fs by propagating the initial wavefunction through vacuum for about 2 ps. The resulting diffraction pattern is shown in figure 4(d). Compared with the bandwidth-limited result, the individual photon orders in $k_x$ are more washed out. However, a pronounced deflection of more than $1^{\circ}$ remains visible, which appears to be readily detectable with available UPEM setups. The physical principles underlying the generation of the diffraction patterns are very similar to those discussed above.

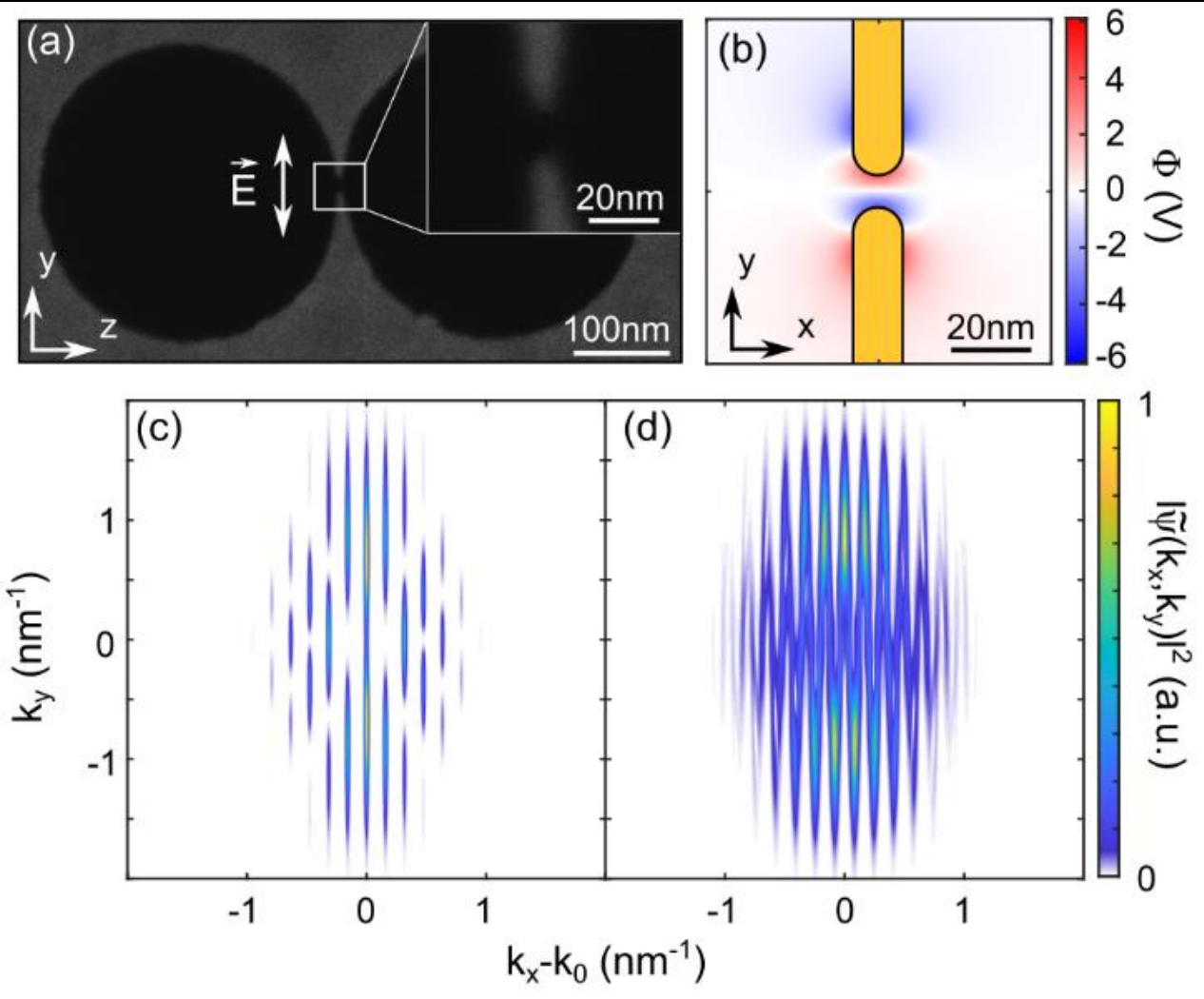

Figure 4. Simulation of near-field low-energy electron diffraction by a realistic sample geometry. (a) SEM image of a nanoresonator milled into a 13-nm-thick free-standing gold film. The gap is 10 nm wide (see inset). (b) Near field potential around the optically excited structure, approximated by dipole distributions on both tips of the nanoresonator. Typical field enhancements of such gap antennas are above 20, suggesting that such potentials can be reached with incident amplitudes $E_L$ below 0.03 V/nm. (c) Final momentum density for an incident electron wavepacket with 20 fs time resolution, 100 eV energy and 6 nm transverse spread. The transverse spread is chosen to let the electron pass through the nanogap. The energetic width of this bandwidth-limited wavepacket is 0.05 eV. (d) Same as in (c) but for a chirped wavepacket with a kinetic energy spread of 2 eV. Both simulations reveal a pronounced angular deflection of the electron by the localized near-field interaction.

## 4. Conclusion

We analyzed the 2D diffraction of nonrelativistic electron wavepackets by the optical near field potential of individual, small nanostructures. For this, we have performed analytical and numercial calculations of the 2D schrödinger equation. Their solutions show rich diffraction patterns in momentum space. In the direction along the electron propagation, we observe the well-known PINEM sidebands at multiples of the photon energy. Additionally, a diffraction pattern is seen in the transverse direction, which becomes even more pronounced for slow electrons with ~100 eV kinetic energies. This modulation of the wavepacket at the specific photon orders is defined by the transverse variation of the near field potential at the corresponding longitudinal Fourier components. Higher photon-order interactions emerge through higher powers of the transverse potential variation and result in stronger structured transverse diffraction patterns.

The analytical calculations allow for an efficient study of the effect of experimental parameter variation on the diffraction pattern like, for example, the electron kinetic energy, structure size, the shape of the potential, or the electric field strength. Specifically for slow electrons, we identify the conditions for optimum coupling. For nanostructures in the 10-nm range, the simulations predict efficient scattering with wide-angle angular deflection patterns that appear well resolvable in existing ultrafast low-energy electron microscopes. Making use of the intrinsic high temporal resultion of UPEM, this opens the way to using slow electrons for a full vectorial characterization of the dynamics of transient, localized near fields around single nanostructures.

Acknowledgements

We thank the Deutsche Forschungsgemeinschaft for support within the priority program QUTIF (SPP1840). Additional support from SPP1839 and the Volkwagen-Stiftung (SMART) is acknowledged. We have performed simulations at the HPC Cluster CARL in Oldenburg (DFG INST 184/157-1 FUGG).